\documentclass[%
 reprint,
 amsmath,amssymb,
 aps,
prstab,
showkeys,
superscriptaddress
]{revtex4-1}
\pdfoutput=1
\usepackage[normalem]{ulem}
\usepackage{hyperref}
\usepackage{gensymb}
\hypersetup{
    colorlinks=true,
    linkcolor=blue,
    filecolor=magenta,      
    urlcolor=blue,
    citecolor = blue,
    bookmarks=false,
}
\usepackage{hhline}
\usepackage{graphicx}
\usepackage{dcolumn}
\usepackage{bm}

\newcommand{\subsc}[1]{\ensuremath{_{\textrm{\begin{scriptsize}#1\end{scriptsize}}}}}
\newcommand{\supsc}[1]{\ensuremath{^{\textrm{\begin{scriptsize}#1\end{scriptsize}}}}}

\begin{document}

\title{Coaxial multi-mode cavities for fundamental superconducting rf research in an unprecedented parameter space}


\author{P. Kolb}
 \email[Email: ]{kolb@triumf.ca}
 \affiliation{TRIUMF, 4004 Wesbrook Mall, Vancouver, B.C. V6T 2A3, Canada}
\author{Z. Yao}
 \affiliation{TRIUMF, 4004 Wesbrook Mall, Vancouver, B.C. V6T 2A3, Canada}
\author{T. Junginger}
 \affiliation{TRIUMF, 4004 Wesbrook Mall, Vancouver, B.C. V6T 2A3, Canada}
 \affiliation{University of Victoria, Victoria, B.C., Canada}
 \author{B. Dury}
 \affiliation{University of British Columbia, Vancouver, B.C., Canada}
\author{A. Fothergill}
 \affiliation{University of British Columbia, Vancouver, B.C., Canada}
 \author{M. Vanderbanck}
 \affiliation{University of British Columbia, Vancouver, B.C., Canada}
 \author{R.E. Laxdal}
 \affiliation{TRIUMF, 4004 Wesbrook Mall, Vancouver, B.C. V6T 2A3, Canada}
\date{\today}	

\begin{abstract}
 Recent developments in superconducting radio-frequency (SRF) research have focused primarily on high frequency elliptical cavities for electron accelerators. Advances have been made in both reducing RF surface resistance and pushing the readily achievable accelerating gradient by using novel SRF cavity treatments including surface processing, custom heat treatments, and flux expulsion. Despite the global demand for SRF based hadron accelerators, the advancement of TEM mode cavities has lagged behind. To address this, two purpose-built research cavities, one quarter-wave and one half-wave resonator, have been designed and built to allow characterization of TEM-mode cavities with standard and novel surface treatments. The cavities are intended as the TEM mode equivalent to the 1.3\,GHz single cell cavity, which is the essential tool for high frequency cavity research. Given their coaxial structure, the cavities allow testing at the fundamental mode and higher harmonics, giving unique insight into the role of RF frequency on fundamental loss mechanisms from intrinsic and extrinsic sources. In this paper, the cavities and testing infrastructure are described and the first performance measurements of both cavities are presented. $R_s(T)$ data is analysed to extract both the temperature dependent, $R_{Td}$, and temperature independent, $R_{Ti}$, components of the surface resistance and their dependence on RF field and frequency. In particular, the $R_{Td}$ was found to be at low fields $\propto \omega^{1.9(1)}$ at 4.2\,K and $\propto \omega^{1.80(7)}$ at 2\,K, fairly well with the theoretical model. The growth of $R_{Td}$ with increasing field amplitude matches both exponential and quadratic growth models fairly well in the examined range of $B_p$. $R_{Ti}$ is determined to be $\propto \omega^{\sim 0.6}$, matching roughly with anomalous losses, while no clear field dependence was determined. In addition, first measurements of a 120\degree\,C baking treatment and of the external magnetic field sensitivity are presented. 
\\
\end{abstract}
\maketitle
\section{Introduction}
Nuclear physics experiments rely on superconducting radio-frequency (SRF) heavy ion particle accelerators such as the ISAC-II \cite{ISAC}  facility at TRIUMF to study the nuclear structure of rare isotopes among other topics of research. New large driver accelerators for hadron facilities such as FRIB \cite{Leitner:SRF2013-MOIOA01}, RAON \cite{Jeon:SRF2015-MOAA05},  PIP-II \cite{Lebedev:2015uuu}, ESS \cite{Peggs:2012ESS, Schlander:SRF2017-MOYA01}, and C-ADS \cite{Chi:LINAC-TU1A03} are being installed or developed to support a variety of research interests. To increase the energy of the beam in the velocity regime up to $\beta \leq 0.6$, these accelerators use different types of  TEM-mode SRF cavities, such as quarter-wave resonators (QWR) and half-wave resonators (HWR) at frequencies ranging from 80 to 400\,MHz. SRF research is essential to advance particle accelerator technology. Higher gradients result in shorter, more economical linear accelerators (LINACs) or higher energies for the same accelerator length. As these SRF cavities are typically cooled with liquid helium at temperatures near 2.0 or 4.2\,K, the RF losses in the cavity walls are a major cost driver in capital investment and in operating budget for the cryoplant and its infrastructure. Higher quality factors $Q_0$ mean smaller cryoplants can be used for the same amount of accelerating voltage. Despite the strong interest in TEM mode cavities for new hadron projects, the bulk of the recent developments to enhance cavity performance have been performed on 1.3\,GHz $\beta = 1$ cavities in support of existing projects such as EU-XFEL \cite{XFEL}, LCLS-II \cite{Galayda:IPAC2018-MOYGB2}, and future projects such as the ILC \cite{adolphsen2013international, bambade2019international}. Advances have been made in both enhancing the quality factor $Q_0$, which corresponds to a lowered surface resistance $R_s$, but also in pushing the readily achievable accelerating gradient $E_{acc}$ to higher levels by using novel SRF cavity treatments with a focus on improved surface processing, customized heat treatments, and a better understanding of flux expulsion \cite{Ciovati:120Cbake, grassellino:Degassing, Reschke:SRF07-TUP, grassellino2018:TwoStepBake,Grassellino:Infusion, Grassellino:Doping}. Systematic studies have not been reported on low frequency, low $\beta$ TEM-mode cavities.\\
Much of the research for 1.3\,GHz applications has been done on single-cell cavities. These are compact cavities, not intended for acceleration, but designed with similar features such as RF frequency, peak surface field to accelerating gradient ratios and identical accelerating mode to the typical 9-cell variants designed for on-line acceleration. Single cell cavities are relatively inexpensive and have been duplicated around the world to allow treatment comparison between research centers throughout the SRF community and greatly enhance development progress.\\
For TEM mode cavities such a focused global development is much more difficult since the design space is broader in terms of hadron velocity and RF frequency. Projects typically optimize cavity parameters within the project and design a few unique cavity designs to span the intended velocity range of a particular LINAC. A coaxial test cavity analogous to the 1.3\,GHz single cell cavity would serve to shift SRF research away from project driven design to a more focused study of the TEM geometry and frequency range and offer a systematic way to enhance cavity performance. In addition the coaxial geometries can be tested at not only the fundamental eigenmode but also at higher harmonics enabling data sets at several RF frequencies within the same cooldown cycle and for the same cavity treatment, surface roughness, RRR  and environmental conditions.\\
This paper reports on the design and first performance characterizations of two coaxial test cavities; a quarter wave resonator and a half wave resonator. The cavity geometries represent the two main structure types used in hadron LINACs to date, namely a QWR and a HWR.. Each cavity is designed to operate in the fundamental and several higher order modes with similar RF characteristics in terms of peak surface field ratio $E_p$/$B_p$. The two resonators are intended to be used for a broad array of fundamental studies. These studies includes the measurement of RF surface resistance as a function of peak magnetic surface field $B_p$ and temperature $T$, and the sensitivity of the geometries to trapped magnetic flux, all as a function of RF frequency and different cavity treatments. The design and implementation of the cavities will be presented, as well as first results.\\
This paper is structured as follows: Section \ref{Sec_motivation} motivates the development of the presented SRF cavities and tools.  Section \ref{sec_methodology} will describe the cavity design and goes over details of the surface preparation, testing methodology, and available tools. Section \ref{secDiscussion} shows the cavity performance as function of peak surface field $B_p$ for a conventional surface treatment. Also, $Q_0(T)$ data collected during the cooldown from 4\,K to 2\,K is analysed. In addition, performance measurements of the QWR after 120\degree\,C baking are presented, as well as flux sensitivity data is shown for the QWR as an example of characterizations of flux expulsion from TEM mode cavities. Section \ref{secConclusions} presents a summary of the presented work, including an outlook into future work.
\begin{figure}
\centering
\includegraphics[width=0.9\linewidth]{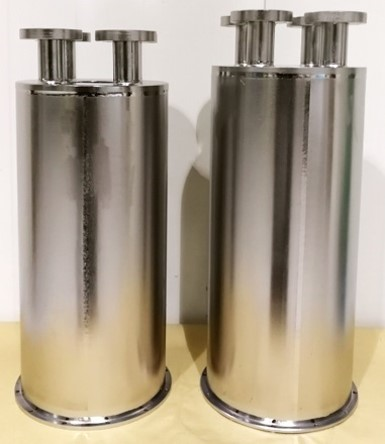}
\caption{\label{Fig_Cavities} Photo of the two coaxial cavities.}
\end{figure}\\
\section{Motivation\label{Sec_motivation}}
\begin{figure}
\centering
\includegraphics[width=0.9\linewidth]{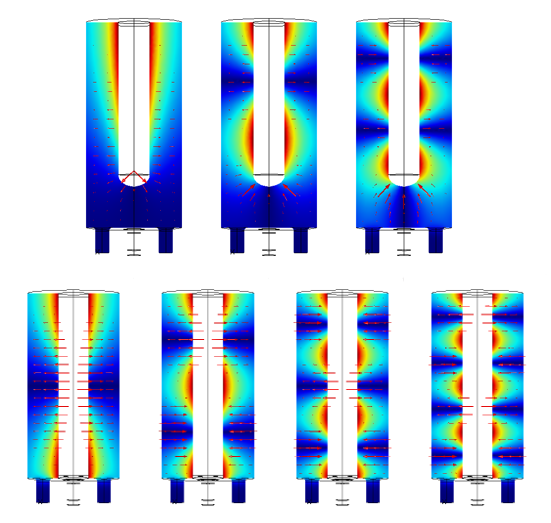}
\caption{\label{Fig_CavitiesFieldDistri} Field distribution in the QWR (top) and HWR (bottom) in the investigated eigenmodes. The heat-map shows the magnetic field and the arrows the electric field.}
\end{figure}

More and more hadron LINACs using SRF technology are being designed and constructed as centerpieces of facilities such as FRIB, RAON, PIP-II, ESS, and C-ADS. Despite this trend, a systematic analysis of the surface resistance in TEM mode cavities globally has not been undertaken due to the broad parameter space in choosing cavity types, geometric $\beta = v/c$ values, and RF frequencies. Heat treatments developed on 1.3\,GHz single cell cavities and rolled out on production nine-cell units have not been employed on TEM mode cavities except for degassing at 650\degree - 800\degree\,C and the 120\degree\,C in-situ vacuum bake \cite{Ferdinand:LINAC-MO201}. Flux expulsion studies, that were instrumental at understanding how to achieve the highest quality factors in continuous wave (cw) 1.3\,GHz applications, have not been systematically undertaken. \\
Several open questions remain concerning the performance of TEM mode cavities. What is the source of the medium field Q-slope at 4\,K that has forced some projects to choose operation at 2\,K over 4.2\,K, despite the reduced losses that come with low frequency and added technical complexity of 2K operation?  What customized heating or doping treatments optimized for 1.3\,GHz would help to lower the surface resistance $R_{s}$ at 4\,K for low frequency TEM mode cavities? Is there a flux expulsion technique that would benefit TEM mode cavities to lower the residual resistance?\\
Using a dedicated purpose built set of coaxial cavities allows tackling these questions, advancing the understanding of TEM mode cavities, and shedding light on the role of the RF frequency in cavity performance in a systematic way. \\
Two cavities were designed; one QWR and one HWR. The QWR has a fundamental resonance frequency of 217\,MHz and the HWR has a fundamental resonance frequency of 389\,MHz. The fundamental RF frequencies of the cavities were chosen to be as low as possible to cover commonly used frequencies, and at the same time fit in a pre-existing induction furnace sized for 1.3GHz single cell cavities, to allow customized heat treatments. Both cavities are shown in Fig. \ref{Fig_Cavities}. The performance of these cavities is characterized via measurement of $Q_0$ as function of the RF field amplitude expressed in the form of the peak surface fields $E_p$ and $B_p$ not only in their fundamental eigenmode, but also their higher order modes (HOMs) to determine the dependence of the cavity performance on frequency without changing the cavity or environmental influences. The field distribution of the fundamental mode and HOMs of interest for the two cavities are shown in Fig. \ref{Fig_CavitiesFieldDistri}. Multi-mode performance characterization allows for an expansion of the parameter space in terms of frequency and field amplitude. Combined with the available parameter space in temperature, external magnetic field, and surface treatment, a previously unavailable parameter space is now available without changing intrinsic and extrinsic factors to the cavity. To fully explore this parameter space, several infrastructure developments were done at TRIUMF to be able to determine the dependence of the surface resistance on all before mentioned factors. \\
A critical part of understanding the SRF cavity performance is the temperature dependence of the surface resistance $R_s$, which can be expressed as 
\begin{align}
R_s(T) = R_{Td}(T) + R_{Ti},
\end{align} 
with $R_{Td}$ as temperature dependent term and $R_{Ti}$ as temperature independent term. $R_{Td}$ can be calculated numerically based on the Mattis-Bardeen theory \cite{Turneaure} and is approximated \cite{Gurevich_2017} as
\begin{align}
R_{Td}(T) \simeq \frac{\mu_0^2 \omega^2 \lambda^3 \Delta}{\rho_s k_B T}\ln \left(\frac{C_1 k_B T}{\hbar \omega}\right) \exp \left(\frac{-\Delta}{k_B T}\right)   \label{eq_RTd_theory}
\end{align}
with $\omega$ as resonance frequency, $\lambda$ as London-penetration depth, $\Delta$ as energy gap, $\rho_s$ as normal state conductivity, $C_1 \approx 9/2$  and $T$ as temperature. Assuming that these parameters are not frequency dependent, Eq. \ref{eq_RTd_theory} predicts a frequency dependence of $R_{Td}$ $\propto \omega^{\sim1.87}$. \\
One specific goal of the in this paper discussed cavities is the determination of the frequency dependence of the surface resistance and investigate any deviations from theory. Previous research has been done with lead on copper cavities \cite{Szecsi} at low power levels. Other studies have used several elliptical cavities of the same shape but different sizes \cite{Martinello:SRF2017-TUYAA02}. Here the challenge is to ensure that the surface and environmental conditions are comparable for the different test cavities.\\ 
Another approach is to use a sample cavity which can be excited at multiple frequencies such as the Quadrupole Resonator (QPR). The QPR has originally been designed for measurements at 400\,MHz \cite{Mahner}, it was later refurbished for multiple frequencies \cite{JungingerQPR} and optimized by HZB \cite{Kleindienst:SRF2015-WEA1A04}. There are still open questions how to translate results from the QPR to accelerating cavity performance.\\
A HWR type cavity similar to the cavities described here has been developed at Center for Accelerator Science at Old Dominion University\cite{Park:SRF2015-MOPB003, park:srf2019-tufub7}.\\
While Eq. \ref{eq_RTd_theory} explicitly shows a frequency dependence of $R_{Td}$, the field dependence of the overall surface resistance $R_s$ and its components is topic of active research. Several models can describe a commonly observed increase of the surface resistance with applied field. For example pair-breaking \cite{GUREVICH2006}, thermal feedback \cite{GUREVICH2006_02}, and impurity scattering \cite{BAUER200651} models or the so called percolation model \cite{PhysRevSTAB.14.101002} predict a $R_s(B_p) \propto B_p^2$ dependence, while other models, for example a weak superconducting layer on top of the bulk material \cite{Palmieri:SRF03-TUA02} suggest $R_s(B_p) \propto \exp B_p$ dependence. Another non-linear model \cite{PhysRevB.100.064522} tries to include the decreasing surface resistance with increasing RF field, which is observed in nitrogen doped cavities \cite{Grassellino:Doping}.\\
\section{Methodology \label{sec_methodology}}
\subsection{Cavities}
\begin{figure}
\centering
\includegraphics[width=0.9\linewidth]{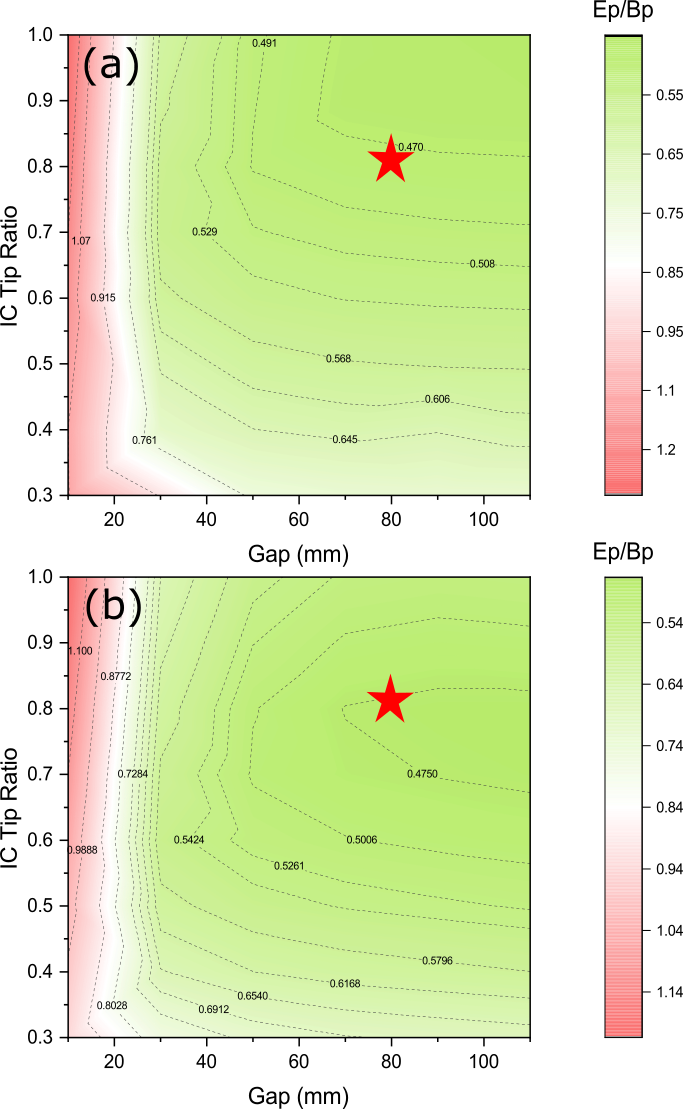}
\caption{\label{Fig_QWR_optim} Heatmaps of $E_p/B_p$ as function of capacitive gap and IC tip ratio to determine a design with a similar $E_p/B_p$ in both the first (a) and second (b) TEM mode. The star marker indicates the chosen design with $E_p/B_p \approx 0.47$\,(MV/m)/mT.}
\end{figure}

The two cavities are used in a similar way as 1.3\,GHz single cell cavities, as pure test cavities in a bath cryostat. To avoid perturbations of the TEM-mode field configuration, beam ports have been removed and all RF ports have been moved to one end plate of the cavities. This is possible as the cavities will not be used for beam acceleration. Since the cavity will be submerged in liquid helium in a bath cryostat, a helium jacket is not necessary. A high shunt impedance and low surface field ratios were not design goals as they would be in an accelerating cavity. Instead the design focused on achieving similar peak surface field ratios $E_p/B_p$ for all the relevant modes as well as the usage of common components such as identically dimensioned outer and inner conductors, identical rinse ports, and the same mechanical components and fixtures. One design limitation was imposed by the size of the induction furnace, which was designed for 1.3\,GHz single cell cavities. This determined that the maximum outer dimensions of the cavities are restricted to a diameter of 200\,mm and a length of 490\,mm. Based on these restrictions, the lowest frequency for the fundamental mode of the QWR was 153\,MHz, regardless of the gap between inner conductor and the bottom plate. A choice was made for the fundamental QWR frequency to be around 200\,MHz and for the HWR to be around 400\,MHz. \\
A straight inner conductor (IC) was chosen to mitigate field distortion in HOMs and at the same time to simplify fabrication. This also allows for a moving T-mapping system to be inserted into the inner conductor. The diameters of the inner and outer conductor were chosen to be 60\,mm and 180\,mm respectively, matching the ISAC-II QWR cavities \cite{Laxdal:pac2001-FPAH111}, allowing for reuse of forming dies. The top and bottom plates are flat, eliminating higher level multi-pacting barriers and simplifying fabrication.\\
Further design choices were made to minimize the peak field ratio E\subsc{p}/B\subsc{p} to push potential field emission onset to higher B\subsc{p} values. \\
For a HWR type cavity of coaxial length $L$ with constant inner and outer conductor radii $a$ and $b$ and peak RF current and voltage of $I_0$ and $V_T$ respectively, the radial electric field $E_r$ and the tangential magnetic field $B_\theta$ at $a \leq r \leq b$ and $ 0 \leq z \leq L$ are given by
\begin{align}
E_r &= -j \frac{\eta I_0}{\pi r} \sin \left(\frac{p\pi z}{L}\right)e^{j\omega t} \label{eq_Er}\\
B_\theta &= \frac{\mu_0 I_0}{\pi r} \cos \left(\frac{p\pi z}{L}\right)e^{j\omega t} \label{eq_Btheta}
\end{align}
with $\omega = p\pi c/L$ with $p = 1,2,3,\dots$, and $\eta = \sqrt{\mu_0/\varepsilon_0}$. Using
\begin{align}
V_T = \eta \frac{I_0}{\pi}\ln\left(\frac{b}{a}\right),
\end{align}
from Eqs. (\ref{eq_Btheta}) and (\ref{eq_Er}) follows that the peak fields $E_p$, $B_p$, and their ratio are
\begin{align}
E_p &= \frac{V_T}{a}\frac{\pi}{\ln\left(\frac{a}{b}\right)}\\
B_p &= \frac{V_T}{ac}\frac{\pi}{\ln\left(\frac{a}{b}\right)}\\
\Rightarrow \frac{E_p}{B_p} &= c
\end{align}
with $c = 1/\sqrt{\varepsilon_0\mu_0}$ as the speed of light. Attention has to be made for the ports, as sharp edges in areas with high surface currents such as the end plates can enhance the magnetic fields, increasing the peak field ratio. The fillet radius at this edge was optimized to mitigate this field enhancement, resulting in no increase of the peak field ratio as can be seen in Tab. \ref{tabRFparam}.\\
\begin{figure}
\centering
\includegraphics[width=0.4\linewidth]{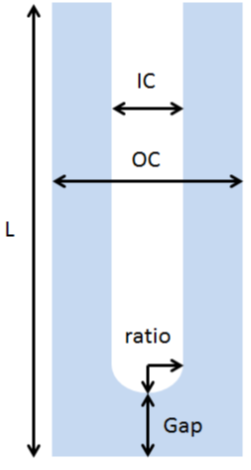}
\caption{\label{Fig_QWR_design} The parametric simulation model of the QWR, in which shaded area shows the RF space in the cavity. Optimization of the peak field ratio focused on the inner conductor tip ratio and gap.}
\end{figure}
For a QWR type cavity, the $E_p$/$B_p$ values are determined by the geometry of the IC tip and are therefore higher than the HWR values. Optimization of the QWR geometry focused on the IC tip cap, described by the ratio of vertical to horizontal size of the tip, and capacitative gap, with the parameter space shown in Fig. \ref{Fig_QWR_design}. The optimization considered both the fundamental mode at 200\,MHz and next higher TEM mode at around 600\,MHz and is shown in Fig. \ref{Fig_QWR_optim}. A peak field ratio of 0.47\,(MV/m)/mT was reached.\\ 
Both cavities are equipped with four cleaning ports for accessing the RF volume with a wand to high pressure rinse (HPR) the cavity. All four ports are on the same flat plate. From each port, the water jet from a nozzle covers about 1/3 of the cavity surface, providing sufficient overlap between rinse ports to cover the whole cavity. To prevent RF losses on non-niobium parts, the rinse ports are 60\,mm long. \\
The cavities are made from pure Niobium to prevent contamination with foreign materials during heat treatments. High residual resistivity ratio (RRR) niobium is used on the main body while the port flanges and QWR bottom plate are made from reactor grade Niobium. Since these components see minimal if any RF fields, the reactor grade niobium can be used without any loss of performance to reduce fabrication costs. Each cavity is a single body with all parts electron-beam welded together without a removable bottom plate. This prevents that the RF field reaches any non-niobium surface, such as vacuum gaskets. Vacuum seals are realized with indium wire seals on the four ports. \\
Further details of the cavity design can be found in \cite{Yao:SRF2017-TUPB065}. Resonant frequencies, as well as numerically calculated peak field ratios $E_p/B_p$ and geometric factors 
\begin{align}
G  = \omega\mu_0 \frac{\int_V |\textbf{H}|^2 dV}{\int_S |\textbf{H}|^2 dS},
\end{align} 
with $\omega$ as resonant frequency and $H$ as the magnetic field, for the TEM modes of interest are listed in Table \ref{tabRFparam}.\\
\begin{table}
\caption{\label{tabRFparam} RF parameters of the two cavities for the TEM modes under investigation.}
\begin{center}
\begin{tabular*}{\linewidth}{l@{\extracolsep{\fill}}rrr}
\hline\hline 
Cavity 	& Frequency 	& E\subsc{p}/B\subsc{p} 	& G			\\	 
		& [MHz] 		&[(MV/m)/mT] 				& [$\Omega$]\\ \hline
QWR 	& 217			& 0.4796					& 37.47 	\\
QWR 	& 648			& 0.4679					& 113.7		\\ \hline
HWR 	& 389			& 0.2975					& 60.39 	\\
HWR 	& 778			& 0.2981					& 120.77	\\
HWR 	& 1166			& 0.2981					& 181.8		\\
HWR 	& 1555			& 0.2990					& 241.24	\\
\hline\hline
\end{tabular*}
\end{center}
\end{table}
\subsection{Available Infrastructure}
A crucial part of the novel cavity treatments is high temperature treatment at in the range from 100\degree to 1000\degree\,C for a specified amount of time in either a ultra high vacuum or low pressure environment. For this, the TRIUMF induction furnace is used. To study the effects of external magnetic fields and flux expulsion, a set of of 3D Helmholtz coils was designed and built around the cavities and existing cryostat. To control the cavities, the existing RF setup with some modifications is used. In this section, this infrastructure is described.
\subsubsection*{Induction Furnace}
\begin{figure}
\centering
\includegraphics[width=0.95\linewidth]{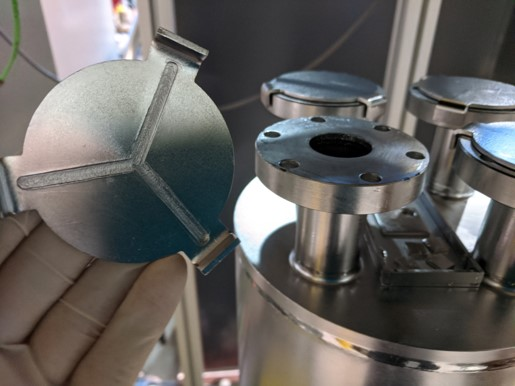}
\caption{\label{fig_IndFurCaps} Slotted Nb caps used to limit line of sight towards the RF surface while providing a defined and reproducible leak for gas flow from RF volume to the furnace volume.}
\end{figure}

\begin{figure}
\centering
\includegraphics[width=0.95\linewidth]{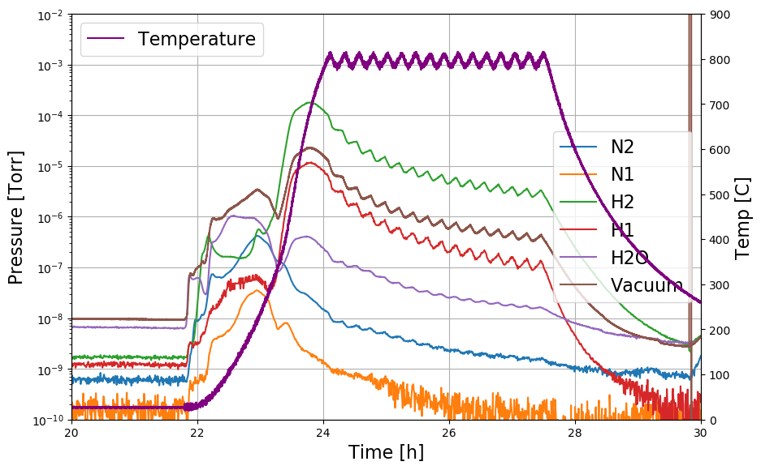}
\caption{\label{fig_IndFurRGA} Sample degassing treatment of the HWR at 800\degree\,C. Shown are also partial pressures of gasses of interest, showing a strong removal of hydrogen from the cavity.}
\end{figure}

For high temperature heat treatments such as degassing \cite{grassellino:Degassing}, nitrogen-doping \cite{Grassellino:Doping}, or nitrogen-infusion \cite{Grassellino:Infusion}, the TRIUMF induction furnace is used. The design is based on the JLab induction furnace \cite{Dhakal:IndFurnace} and dedicated to be used only for Nb SRF cavities. In this furnace, a niobium susceptor is heated via RF induction. The heat generated in the susceptor is transferred to the cavity via radiation. Conventional ultra high vacuum (UHV) furnaces pose a potential contamination risk, requiring the use of caps on the cavity ports \cite{grassellino:Degassing}. In the induction furnace, the RF surface of the cavity has line of sight to only Nb surfaces by design, reducing risk of contamination. In addition, slotted Nb caps (shown in Fig. \ref{fig_IndFurCaps}) placed on the ports of the cavity provide additional line of sight cover while allowing gas flow with a defined and reproducible leak between the UHV space and RF volume of the cavity. An advantage of the caps lies in the reduced effort to clean and refresh the surface via BCP, compared to removing and etching the susceptor of the furnace. A residual gas analyser provides data during the degassing. A sample degassing spectrum during the 800\degree\,C treatment of the HWR is shown in Fig. \ref{fig_IndFurRGA} along with the temperature profile.\\
\subsubsection*{Helmholtz Coils}
\begin{figure}
\centering
\includegraphics[width=0.95\linewidth]{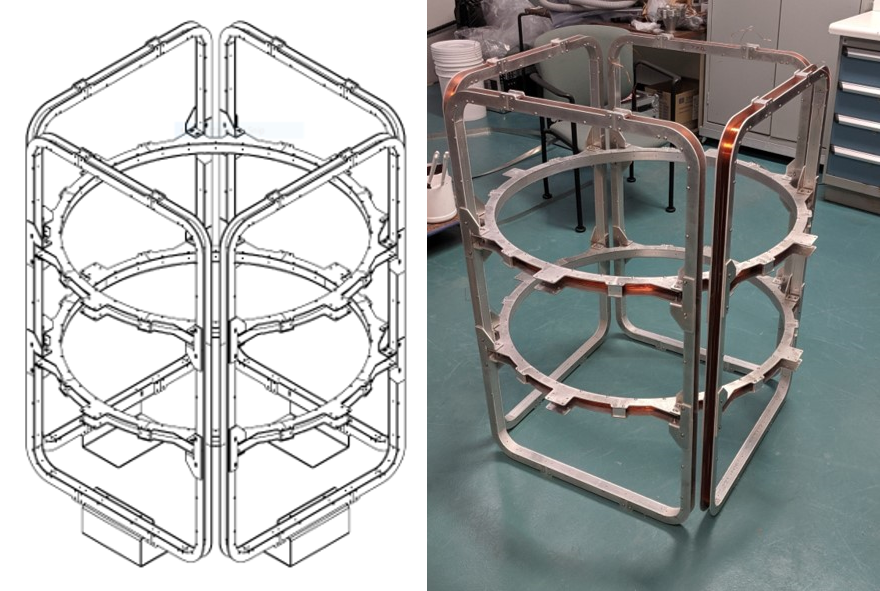}
\caption{\label{fig_HHcoils} Design (left) and realization (right) of the 3D Helmholtz coils before installation in the cryostat.}
\end{figure}

Flux trapping and the expulsion of external magnetic fields can be detrimental to the SRF cavity performance \cite{PhysRevSTAB.16.102002}. For example, the high $Q_0$ performance of nitrogen-doped cavities is very sensitive to external magnetic fields, so much that the specifications in the LCLS-II cryomodules calls for no more than 5\,mG of background field to preserve the high $Q_0$ of the cavities \cite{wu2018achievement}. \\
To control and manipulate the external magnetic field around the cavity in the TRIUMF cryostat, a set of 3 pairs of Helmholtz coils has been designed and built, shown in Fig. \ref{fig_HHcoils}. These coils can be used to either cancel or control the external field to a specific value in all three spacial orientations. The current in each pair of coils can be controlled independently, to allow control of the direction of the field. One of the design criteria for the coils was a field uniformity greater than 95\% over the cavity surface.  To measure the magnetic field, three Bartington Mag F \cite{Bartington} cryogenic flux-gate probes are used. Magnetic field data as well as corresponding temperature data is collected via a Labview \cite{Labview} program. This setup allows studies of how the performance of the coaxial cavities changes under different external magnetic field configurations and cooldown characteristics. \\
\subsubsection*{Chemical Treatment /  surface removal}
\begin{figure}
\centering
\includegraphics[width=0.95\linewidth]{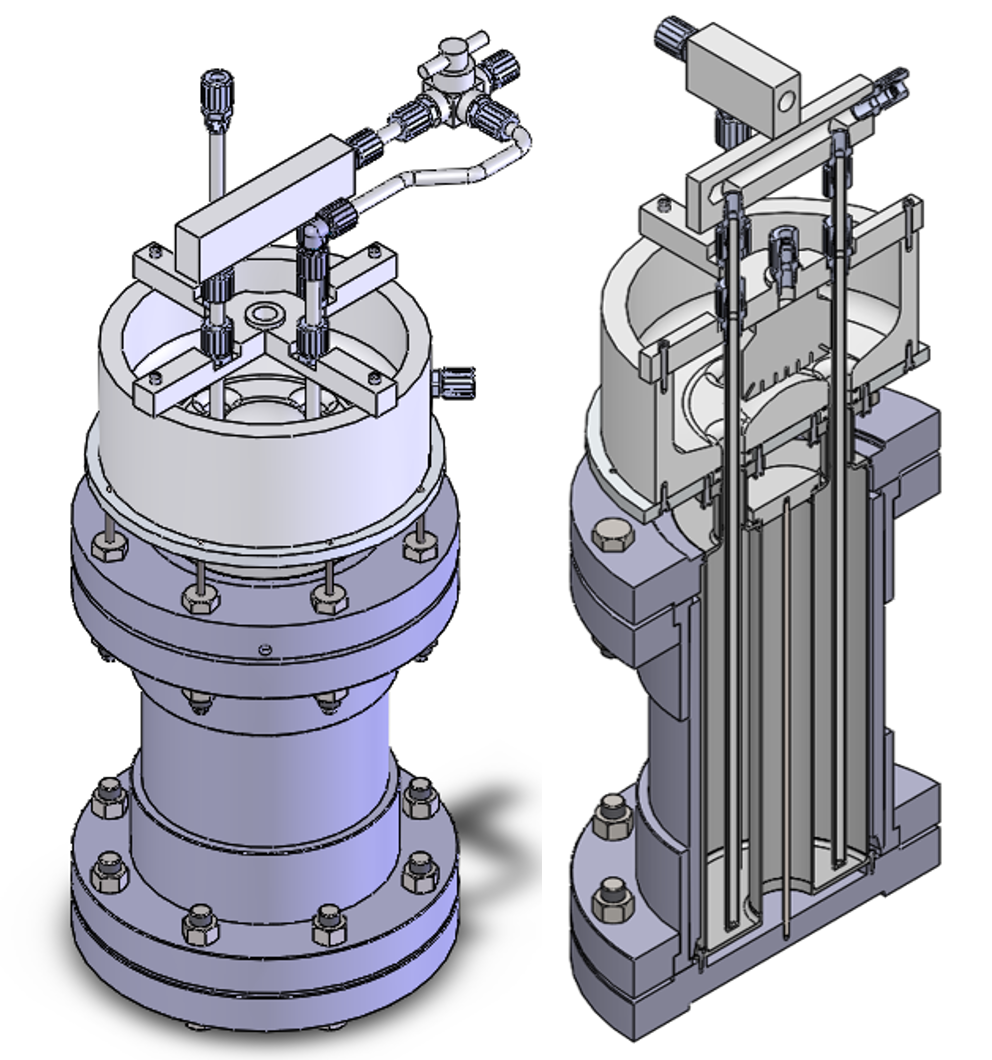}
\caption{\label{fig_BCPsetup} Setup for BCP treatment with water-cooling jacket. Acid is supplied to and pumped out from  the bottom of the cavity via rods through the cavity ports.}
\end{figure}

Chemical surface removal is done via buffered chemical polishing (BCP) in a standard 1:1:2 mixture of hydrofluoric acid, nitric acid,  and phosphoric acid. Fig. \ref{fig_BCPsetup} shows the design for the mechanical setup. Acid is supplied through a manifold and pumped to the bottom of the cavity via a diffuser which prevent fast flowing jets of acid. An overflow reservoir at the top of the cavity ensures that all of the RF surface is in contact with the acid. From the reservoir the acid flows back into the acid storage barrel, ensuring a constant flow of fresh acid through the cavity. The whole cavity is strapped into a water cooling jacket to regulate the cavity temperature. The acid temperature in the storage barrel is controlled with a heat exchanger, which draws from the same cooling water. To cool the water a 7\,kW chiller from Advantage Engendering \cite{AdvantageEngineering} is used. Water temperatures of between 10\degree\,to 12\degree\,C are typically used. This results in etching rates of around 1\,$\mu$m/s. The manifold is also used to pump out the acid and supply the cavity with rinse water once the etch is done.
\subsubsection*{RF Setup}
A self-excited loop (SEL) is utilized in the low level RF (LLRF) control of cryostat cold tests at TRIUMF. The SEL frequency tracks the resonant frequency of the cavity. The frequency control is stable in either open or closed amplitude loop and free of ponderomotive instabilities. SEL, in absence of phase loop feedback, is ideal for cavity performance characteristics, multipacting conditioning and high-power pulse conditioning.\\
The LLRF boards developed for ISAC-II and ARIEL e-Linac projects \cite{Laverty:pac200*-FPAH111} control at 140\,MHz and allow for either pulsed or continuous wave (cw) operation. An intermediate frequency is employed to down-convert the cavity frequency to 140 MHz for input, and to up-convert the output signal to the cavity resonant frequency for driving the RF amplifier. One essential part of the frequency converter is the high-performance bandpass filter. Discrete filters with  $<$-20\,dBc at ±30\,MHz were chosen for 200\,MHz and 400\,MHz, while cavity filters with $<$-30\,dBc in the same range for the higher frequency modes are used. The intermediate frequency and the bandpass filter are switched when changing test modes.\\ 
Two wide-band solid-state amplifiers from BEXT \cite{BEXT} (70 to 650\,MHz, 500\,W) and R\&K \cite{RK} (650 to 2800\,MHz, 350\,W) are used to cover the frequency spectrum from 70\,MHz to 2.8\,GHz with up to 500\,W of RF power.\\
Two variable RF couplers are used for the two different cavities: one based on an antenna coupler for the QWR to transfer power via the electric field and the other with a loop antenna for the HWR which couples to the magnetic field. $Q_{ext}$ for both couplers varies by 5 orders of magnitude over a travel of 30\,mm, while a maximum travel distance of 40\,mm is available. The couplers provide a large range of $Q_{ext}$ to enable operation at critical coupling for any RF mode.\\
To accurately measure $Q_0$, the coupler is moved to critical coupling. From a decay time measurement at low RF field, coupled with power and frequency measurements, $Q_0$, $Q_{pu}$, and $B_p$ are determined. This calibrates the setup for further measurement in continuous wave operation. Measurement uncertainty in both $Q_0$ and $B_p$ is determined by the remaining mismatch between cavity and coupler $Q$-values during the calibration measurement, expressed as deviation of the standing wave ratio (SWR) from 1. This typically results in relative uncertainties of around 5-10\,\% for $Q_0$ and 2-5\,\% for $B_p$. Other systematic sources of uncertainty like instrument precision of power meters and frequency counters are considerably smaller, and are therefore not considered.
\subsection{Data Preparation and Fitting}
\begin{figure}
\centering
\includegraphics[width=0.95\linewidth]{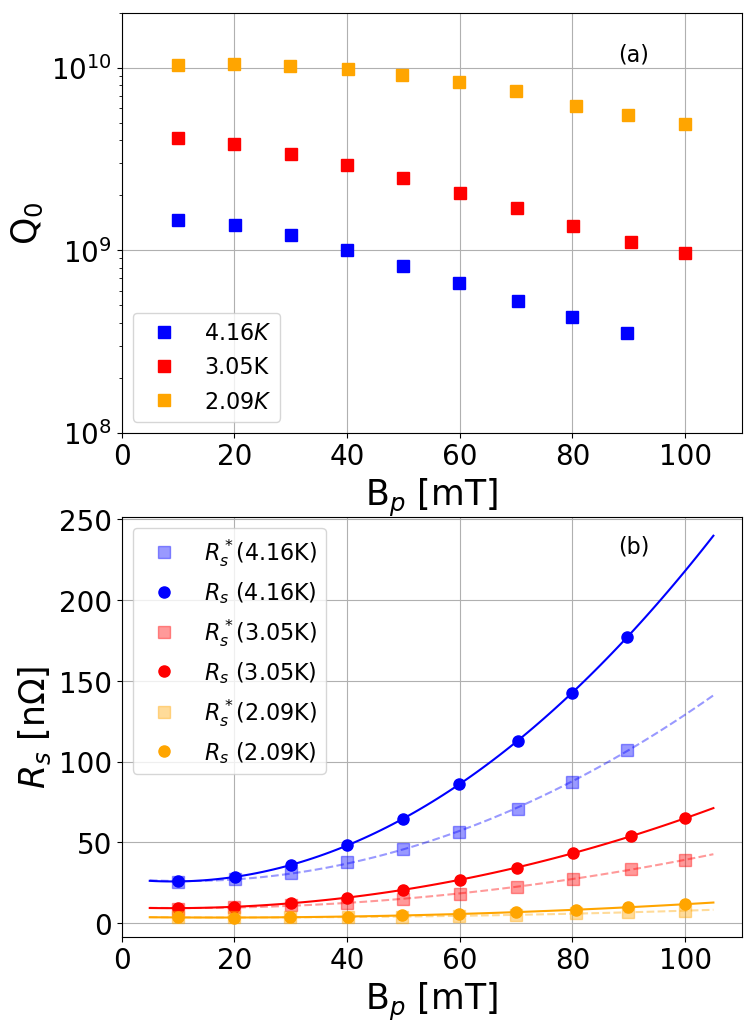}
\caption{\label{Fig_Rscorrection} $Q_0(B_p)$ for three different temperatures, markers in (a), is first converted into $R_s^*$, square markers in (b), using Eq. (\ref{eq_G}). $R_s^*$ is then fitted to Eq. (\ref{eq_Rs*}), dashed lines in (b), which then is corrected into $R_s$, represented by the solid line and circular markers in (b). Error bars are omitted for clarity.}
\end{figure}

\begin{figure}
\centering
\includegraphics[width=0.95\linewidth]{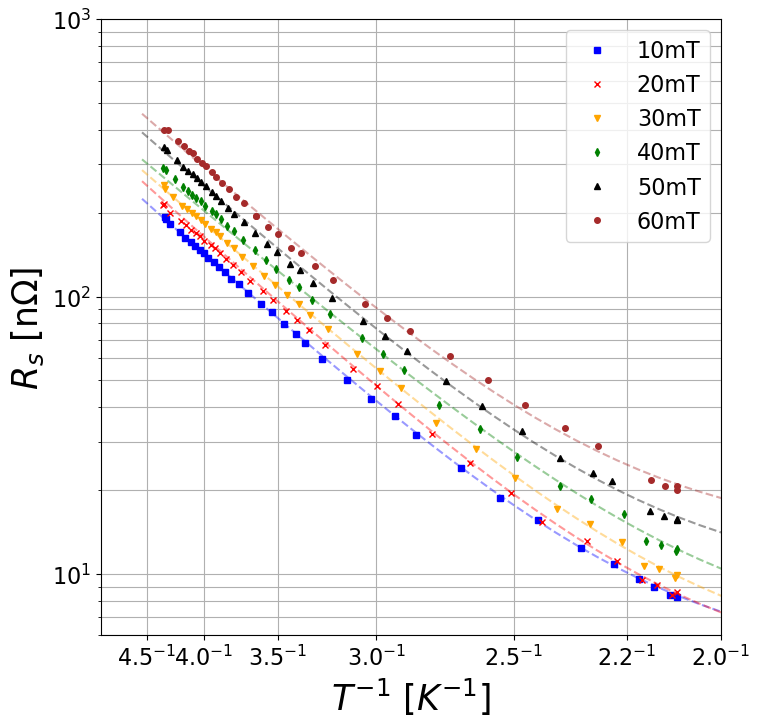}
\caption{\label{fig_Rs_T_fit} Example of the $R_{s}$ fitting in the 648\,MHz mode with data ranging from B\subsc{p} = 10$\dots$60\,mT. The markers represent the measured data, while the dashed lines represent the fit of Eq. (\ref{eq_Rs_T}) to the data. Error bars are omitted for clarity.}
\end{figure}

Initial analysis is done by converting the quality factor data to the average surface resistance, $R_s^*$, through the well known approximation 
\begin{align}
R_s^*=G/Q, \label{eq_G}
\end{align}
where $G$ is the geometry factor defined as $G = \omega\mu_0 \int_V H^2 dV / \int_S H^2 dS$. Field distributions and values for $G$ for all modes have been computed using \textsc{COMSOL} \cite{COMSOL} and are given in Table \ref{tabRFparam}.\\
Due to the non-uniform field distribution over the cavity surface and the field dependence of $Q_0$, the conversion $R_s^* = G/Q_0$ does not reveal the true field dependence of the surface resistance and a correction has to be applied. This correction is especially important in TEM mode cavities, as the fields are significantly less uniform over the cavity surface compared to elliptical cavities. A variety of methods \cite{Weingarten:PhysRevSTAB.14.101002, Maniscalco:IPAC2018-WEPMF042, LONGUEVERGNE201841, Kleindienst:SRF2017-THPB054} can be used to extract the true surface resistance field dependence. In the methodology \cite{PhysRevAccelBeams.21.122001} adopted here, $R_s^*(B_p)$ data is first fitted with a power law series
\begin{align}
R_s^*(B_P) = \sum_{\alpha_i} r_{\alpha_i} B_P^{\alpha_i}, \label{eq_Rs*}
\end{align} 
with $r_{\alpha_i}$  as fit parameter. $\alpha_i$ can be any non-negative real value and be chosen to best fit the data. The coefficients $r_{\alpha_i}$ then are corrected using parameters $\beta_{\alpha_i}$, which are derived from the field distribution over the surface of the cavity, resulting in the surface resistance 
\begin{align}
R_s(B_p) = \sum_{\alpha_i} \beta(\alpha_i) r_{\alpha_i} B_P^{\alpha_i}. \label{Eq_Rscorrection}
\end{align}
For a fairly uniform field distribution over the surface, such as in elliptical cavities, the factors $\beta(\alpha_i)$ are close to unity, while for TEM mode cavities, these factors are significant larger than 1. Values for $\beta(\alpha_i)$ for the QWR and HWR modes have been calculated numerically and are given in Tab. \ref{tab_GeoCorrectionBetas}. Note that the HWR values are consistent for each mode indicating that the field pattern is purely coaxial. The QWR values vary between modes due to the changing field pattern around the tip of the inner conductor. An example of the conversion from $Q_0$ to $R_s^*$ to $R_s$ at three different temperatures is shown in Fig. \ref{Fig_Rscorrection} for the 217\,MHz mode of the QWR.\\
\begin{table}
\caption{\label{tab_GeoCorrectionBetas} $\beta(\alpha_i)$ values for the QWR and HWR investigated modes. }
\begin{center}
\begin{tabular*}{\linewidth}{l@{\extracolsep{\fill}}rrr}
\hline\hline 
$\beta_i$ 		&$\beta(0)$ 	&	$\beta(1)$ 	& $\beta(2)$  \\ \hline
QWR - 217\,MHz 	& 1.0		& 1.432 		& 1.778 \\
QWR - 648\,MHz  & 1.0		& 1.473 		& 1.871 \\
HWR - 389\,MHz 	& 1.0		& 1.463			& 1.857 \\
HWR - 778\,MHz	& 1.0		& 1.461			& 1.857 \\
HWR - 1166\,MHz & 1.0		& 1.463			& 1.862 \\
HWR - 1555\,MHz	& 1.0		& 1.463			& 1.862 \\
\hline\hline 
\end{tabular*}
\end{center}
\end{table}
To extract the temperature dependence of $R_s(B_p)$, $Q_0$ is repeatedly measured during the cooldown from 4.2\,K to 2\,K for a number of fixed peak field amplitudes $B_p$ in 10\,mT intervals up to a maximum field of $B_{max}$. Each ramp up of the RF field up to $B_{max}$ is considered as a set, measured roughly at the same temperature $T$ with differences of around 50\,mK between the first and last measurement point in each set. All sets are converted into $R_s^*$ using Eq. (\ref{eq_G}) and fitted to Eq. (\ref{eq_Rs*}) to extract the parameters $r_{\alpha_i}$. In the investigated case, using a polynomial of second order was determined to be sufficient to describe the field dependence in the range of the available data accurately with very small residuals, well within measurement uncertainty. The parameters $r_{\alpha_i}$ are then multiplied by the corresponding $\beta_i$ to determine $R_s$ at the measured field and temperature. All sets are combined, sorted, and split by field amplitude to create new sets of $R_s(T)$. These are fitted using the \textsc{WinSuperFit} \cite{WinSuperFit} code v1.1 for each value of $B_p$ individually to a parametrized version of Eq. (\ref{eq_RTd_theory}) in form of
\begin{align}
R_s(T) &= \frac{a_0}{T}\ln\left(\frac{4k_B T}{\hbar\omega}\right)\exp{\left(\frac{-a_1(T) T_c}{T}\right)} + a_2 \label{eq_Rs_T}\\
&= R_{Td}(T) + R_{Ti}
\end{align}
with $a_0$, $a_1(T)$, $a_2$ as free fit parameters, and $T_c = 9.25$\,K as the critical temperature. $a_1(T)$ represents the superconducting gap $\Delta$ with its temperature dependence
\begin{align}
\frac{\Delta(T)}{\Delta_{T=0}} = \sqrt{1-\left(\frac{T}{T_c}\right)^4}.
\end{align}
$R_{Td}$ and $R_{Ti}$ are the extracted temperature dependent and independent components of the surface resistance respectively. Fit uncertainties in $a_0$, $a_1$, and $a_2$ are propagated into $R_{Td}$ and $R_{Ti}$. Shifts in $\omega$ during the cooldown from 4.2\,K to 2\,K, which are primarily caused by the pressure and Lorentz-force detuning sensitivity of the cavity, are of small order compared to the frequency, and are therefore ignored for the analysis. A collection of these $R_s(T)$ fits for the QWR 648\,MHz mode at fields up to 60\,mT is shown in Fig. \ref{fig_Rs_T_fit}. This is done for all measured modes to extract not only field dependence, but also frequency dependence of these fit parameters and the derived values of $R_{Td}$ at temperatures of interest. The quality of the fits is generally acceptable with R\supsc{2} vales above 0.99, producing fits well within the determined measurement uncertainty. At higher field amplitudes a distinct step in $R_s$ is observed at the $\lambda$-point of liquid helium of 2.17\,K. This is assumed to be an effect caused by a change in cooling capabilities between the normal and superfluid helium. A thorough analysis of this effect is in progress, but beyond the scope of this paper.\\ 
Further data fitting is done in the \textsc{Origin 2020} suite \cite{Origin}, which directly provides uncertainties for the fit parameters as well as R\supsc{2} values. \\
\section{Results\label{secDiscussion}}
\subsection{Cavity Performance Characterization}
\begin{figure}
\centering
\includegraphics[width=0.95\linewidth]{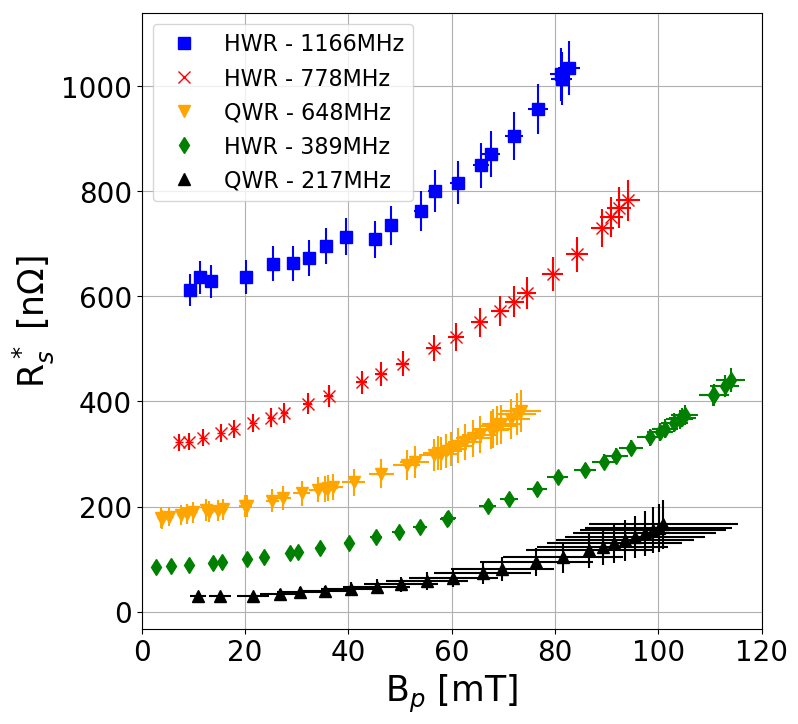}
\caption{\label{fig_RsBp_prebake4K} Measured, uncorrected surface resistance $R_{s}^* (\propto 1/Q_0)$ of the QWR and HWR at 4.2\,K after degassing and 15\,$\mu$m surface removal. The measurement was free of detectable field emissions. The amplitude was limited by quench, except at 1166\,MHz where the amplifier power limit was reached.}
\end{figure}

\begin{figure}
\centering
\includegraphics[width=0.95\linewidth]{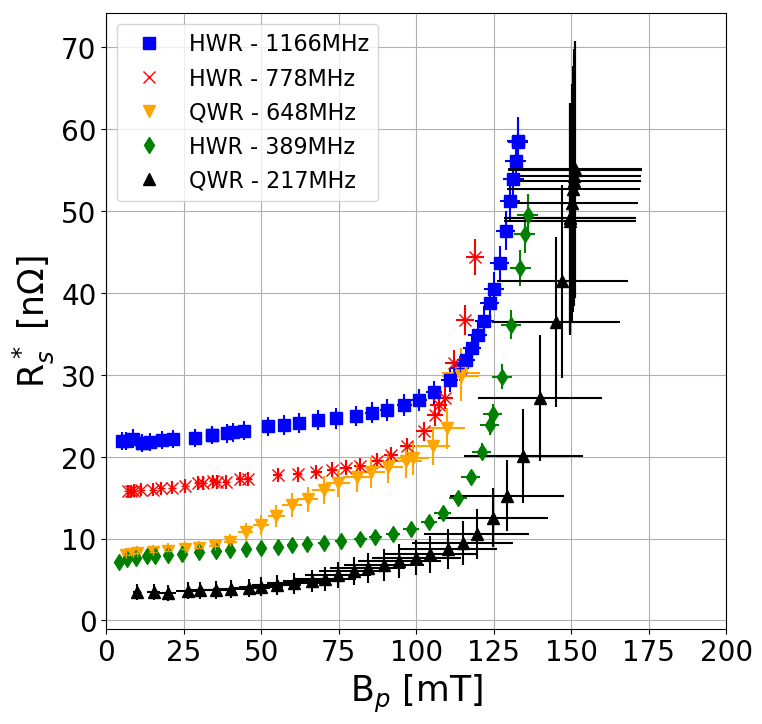}
\caption{\label{fig_RsBp_prebake2K} Measured, uncorrected surface resistance $R_{s}^*$ of the QWR and HWR at 2.1\,K after degassing and 15\,$\mu$m surface removal. The measurement was free of detectable field emissions.}
\end{figure}

The baseline surface treatment for both cavities presented in this paper includes a bulk surface removal via BCP of 120\,$\mu$m, 800\degree\,C degassing in the TRIUMF induction furnace (6\,h for the QWR, 3.5\,h for the HWR; the difference is due to a larger hydrogen content in the QWR) to remove hydrogen from the cavities to prevent Q disease, and a final 15\,$\mu$m BCP surface etch to remove final contaminants. The cavity is then rinsed via high pressure rinsing (HPR) with ultrapure water, dried, and equipped with its pick-up probe, variable coupler, and vacuum connections in a class 10 clean room environment. Initial measurements with the QWR were done in a horizontal orientation (with coaxial axis horizontal). Subsequent QWR tests were done in a vertical orientation. All HWR tests were done in a vertical orientation. Once installed in the cryostat, the quality factor $Q_0$ as a function of peak surface field $B_p$ is characterized at 4.2\,K and 2\,K  at critical coupling with the movable coupler. \\
Combined QWR and HWR performance characterizations of the initial treatment are shown in Fig. \ref{fig_RsBp_prebake4K} for 4\,K and Fig. \ref{fig_RsBp_prebake2K} for 2\,K. The presented data is for the uncorrected surface resistance $R_s^*$. The data for the two QWR modes was collected during a single cooldown of the QWR, same as with the data for the three HWR modes. No field emissions were observed during the measurements, indicating a cavity surface free from particulate contamination. For the presented data, the background field was compensated as close to zero as possible ($< 1$\,$\mu$T) using the Helmholtz coils. \\
In the 4\,K measurements, the surface resistance both increases with increasing field amplitude and with increasing frequency. The overall field dependence follows a similar behaviour in all cavity modes. The field amplitude is limited by quench, except for the 1166\,MHz mode, which is limited by available amplifier power. The QWR has a reduced quench field compared to the HWR due to a different cavity orientation for the initial tests. The horizontal test orientation of the QWR reduced the liquid helium requirement in the dewar, but produced early cw quenches at 4.2\,K due to limited cooling/He-gas buildup in the inner conductor. Maximum quench field in the QWR was 100\,mT ($E_p$ = 47\,MV/m). The HWR reached 115\,mT ($E_p$ =35\,MV/m).\\
At 2\,K, the average surface resistance is decreased significantly compared to the results at 4.2\,K, from 100's of n$\Omega$ to single digit n$\Omega$ in the lowest frequency mode. In medium fields up to 100\,mT, the field dependence of especially the HWR modes seems reduced significantly as well. In the QWR, features in the $R_{s}^*$ curve can be seen at around 60-75\,mT, especially in the 648\,MHz mode. These could indicate insufficiently removed surface contamination after heat treatment. Above 100\,mT peak surface field a strong increase in $R_{s}^*$  is measured without any measured field emission, which is characteristic of high field Q slope (HFQS) \cite{Visetin:SRF03-TUO01}. The quench field was determined to be at 150\,mT (E\subsc{P} = 71\,MV/m) for the QWR and 130\,mT (E\subsc{P} = 40\,MV/m) for the HWR. \\
In the following section, the results of the field distribution corrected $R_s(T)$ fits is presented in terms of the temperature dependent resistance $R_{Td}$ at 4.2 and 2.0\,K, and the temperature independent resistance $R_{Ti}$ based on Eq. (\ref{eq_Rs_T}). These components of $R_s$ are analysed regarding their field and frequency dependence.
\subsection{Temperature Dependent Surface Resistance}
\subsubsection{Field Dependence}
\begin{figure}
\centering
\includegraphics[width=0.95\linewidth]{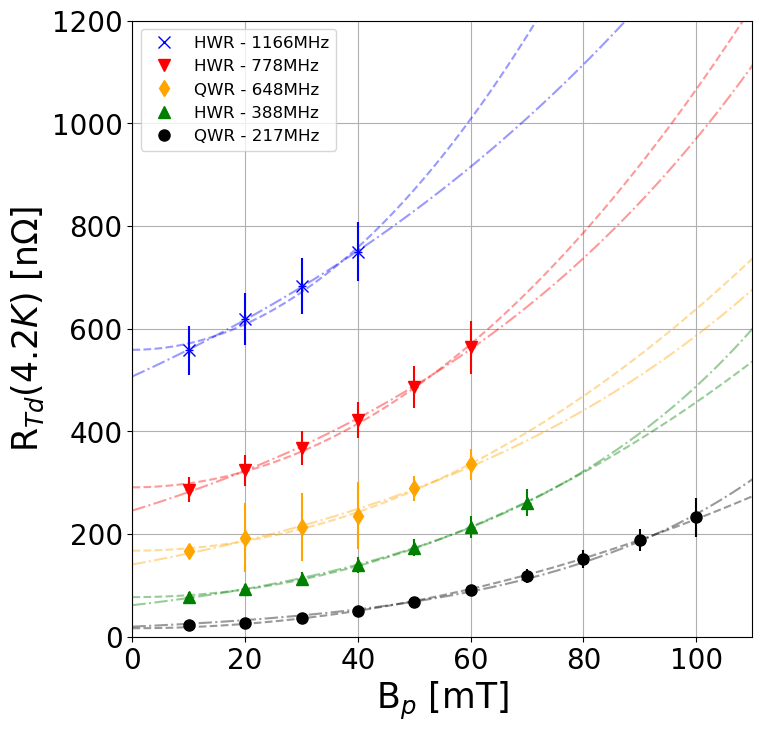}
\caption{\label{fig_RTd4KBp} $R_{Td}$(4.2\,K) as a function of peak surface fields for 5 resonant modes, extracted from the cooldown data. Dashed lines represent fits to the exponential growth, while the dash-dot lines show the quadratic increase. The data fits both trends well.}
\end{figure}

\begin{figure}
\centering
\includegraphics[width=0.95\linewidth]{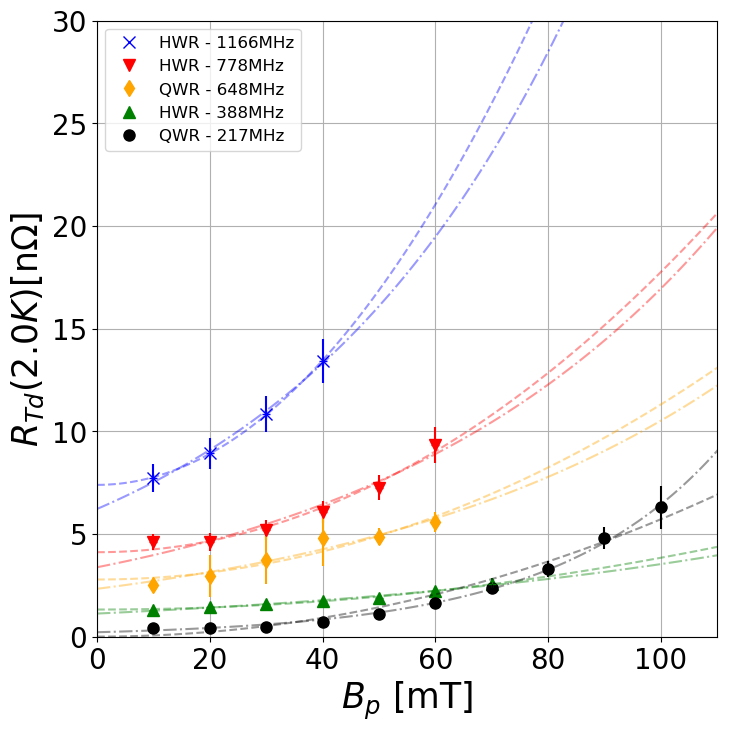}
\caption{\label{fig_RTd2KBp} $R_{Td}$(2\,K) shows a similar, but reduced field dependence compared to $R_{Td}$(4.2\,K). Dashed lines represent fits to the exponential growth, while dash-dot lines show the quadratic increase.}
\end{figure}

Shown in Figs. \ref{fig_RTd4KBp} and \ref{fig_RTd2KBp} are calculated values for temperature dependent component $R_{Td}$ as a function of peak surface field $B_p$ at 4.2\,K and 2\,K respectively. At both temperatures, an accelerated increase of $R_{Td}$ is observed as the RF field increases.\\
The two investigated field dependencies used to describe the increase of $R_{Td}$ are expressed as following: a simple exponential growth
\begin{align}
R_{Td,e}(B_p) =  R_{0,e} \exp\left(\gamma_e \frac{B_p}{B_0}\right)\label{eq_RTd_expGrowth}
\end{align} 
with $R_{0,e}$ as zero-field resistance, $\gamma_e$ as dimensionless growth rate parameter, and $B_0$ as normalizing parameter, which can be freely chosen; and a quadratic increase following 
\begin{align}
R_{Td,q}(B_p) = R_{0,q} \left[1 + \gamma_q \left(\frac{B_p}{B_0}\right)^2\right], \label{eq_RTd_field_Dependence}
\end{align}
with $R_{0,q}$ as zero field resistance and $\gamma_q$ as dimensionless slope parameter.\\
Within the determined uncertainty of $R_{Td}$, both Eqs. (\ref{eq_RTd_expGrowth}) and (\ref{eq_RTd_field_Dependence}) describe the data fairly well as can be seen in Figs. \ref{fig_RTd4KBp} and \ref{fig_RTd2KBp}, where dashed lines represent Eq. (\ref{eq_RTd_expGrowth}) while dash-dot lines visualizes Eq. (\ref{eq_RTd_field_Dependence}). R\supsc{2} values for all fits are above 0.90, with most aggregating above 0.97. Residual differences between the data and the two fit functions are generally of similar magnitude, but slightly lower for the exponential fit. Both describe the data within the determined uncertainties. Thus a definitive statement on the most appropriate field dependence cannot be made by the presented data alone. Supplemental measurements, for example with material science probes such as the $\beta$-SRF beamline \cite{Thoeng:IPAC2018-THPML122} at TRIUMF, would be needed to determine the physics behind the field dependence.\\
At 2\,K, shown in Fig. \ref{fig_RTd2KBp}, $R_{Td}$ is, as expected, significant lower than at 4.2\,K. This is unsurprisingly expressed in a lower zero-field resistance $R_{0}$. Both $\gamma_{e/q}$ parameters on the other hand do not show a clear trend in difference between the two temperatures, indicating that the perceived reduced field dependence at lower temperature is a result of the overall reduced magnitude of the zero-field resistance $R_{0,e/q}$. The fit results for all modes in both temperatures are listed in Tab. \ref{tab_FitParams}.
\begin{table}
\caption{\label{tab_FitParams} Fit parameters for the field dependence of $R_{Td}$, Eqs. (\ref{eq_RTd_expGrowth}) and (\ref{eq_RTd_field_Dependence}),  with $B_0 = 100$\,mT. $R_{0,e/q}$ in n$\Omega$.}
\begin{center}
\begin{tabular*}{\linewidth}{ll|rrrr}
\hline\hline 
Mode 				& T [K]		& $R_{0,e}$ 	&$\gamma_e$ 	& $R_{0,q}$ 	& $\gamma_q$  \\ \hline
QWR - 217\,MHz 		& 4.2		& 16.6(7)		& 2.73(6) 		& 18.7(7) 		& 10.8(6)\\
QWR - 217\,MHz 		& 2			& 0.19(2)		& 3.5(1)		& 0.21(7)		& 20(9)\\
HWR - 389\,MHz 		& 4.2		& 61.5(7) 		& 2.07(2) 		& 76(1)			& 5.1(2)\\
HWR - 389\,MHz 		& 2			& 1.14(4) 		& 1.10(7)		& 1.32(2)		& 1.9(1)\\
QWR - 648\,MHz 		& 4.2		& 144(2)		& 1.40(3)		& 162(2)		& 3.1(1)\\
QWR - 648\,MHz	 	& 2			& 2.18(6)		& 1.59(6)		& 2.48(9)		& 3.7(3)\\
HWR - 778\,MHz 		& 4.2		& 248(2) 		& 1.35(2) 		& 287(5)		& 2.8(2)\\
HWR - 788\,MHz 		& 2			& 3.6(3)		& 1.4(2)		& 4.2(2)		& 3.0(3)\\
HWR - 1166\,MHz 	& 4.2		& 507(3)		& 0.99(2) 		& 557(12) 		& 2.3(3)\\
HWR - 1166\,MHz 	& 2			& 6.3(2)		& 1.9(1)		& 7.38(4)		& 5.2(1)\\
\hline\hline 
\end{tabular*}
\end{center}
\end{table}
\subsubsection{Frequency Dependence}
\begin{figure}
\centering
\includegraphics[width=0.95\linewidth]{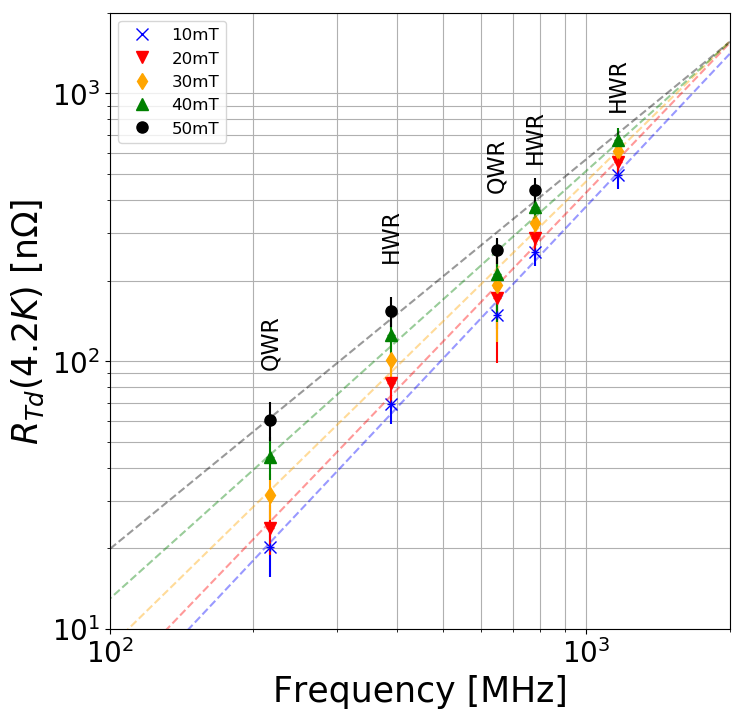}
\caption{\label{fig_RTd4Kfreq} Combined $R_{Td}$ data from QWR and HWR tests at 4.2\,K as a function of frequency for RF fields up to 50\,mT. Dashed lines show best fits of Eq. (\ref{eq_FreqDependence}) to the data, and indicate a decrease of the frequency dependence with increasing RF field.}
\end{figure}

\begin{figure}
\centering
\includegraphics[width=0.95\linewidth]{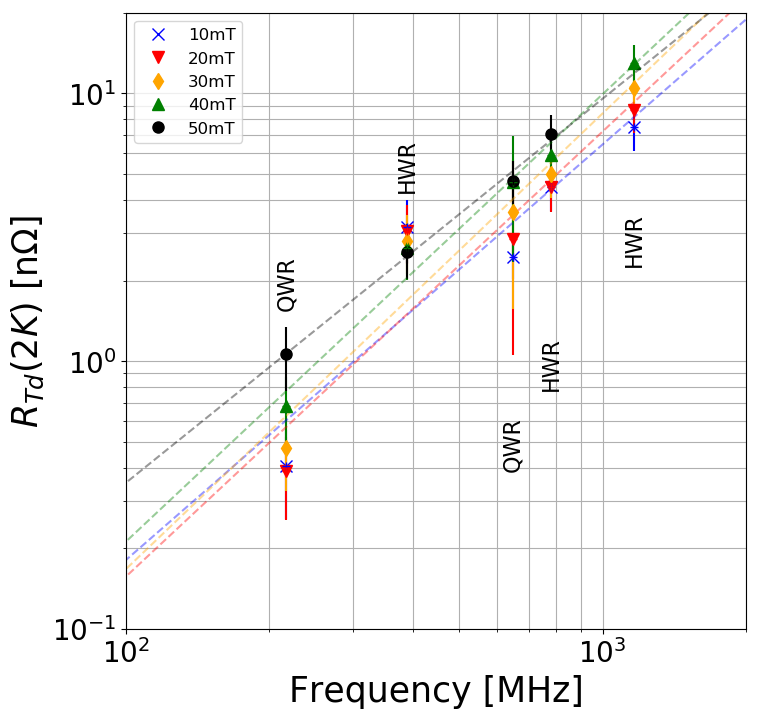}
\caption{\label{fig_RTd2Kfreq} Combined $R_{Td}$ data from QWR and HWR tests $R_{Td}$ at 2\,K as a function of frequency. Dashed lines indicate best fits of Eq. (\ref{eq_FreqDependence}) to the data. A similar trend as at 4.2\,K of a decreasing frequency dependence is observed.}
\end{figure}

\begin{figure}
\centering
\includegraphics[width=0.95\linewidth]{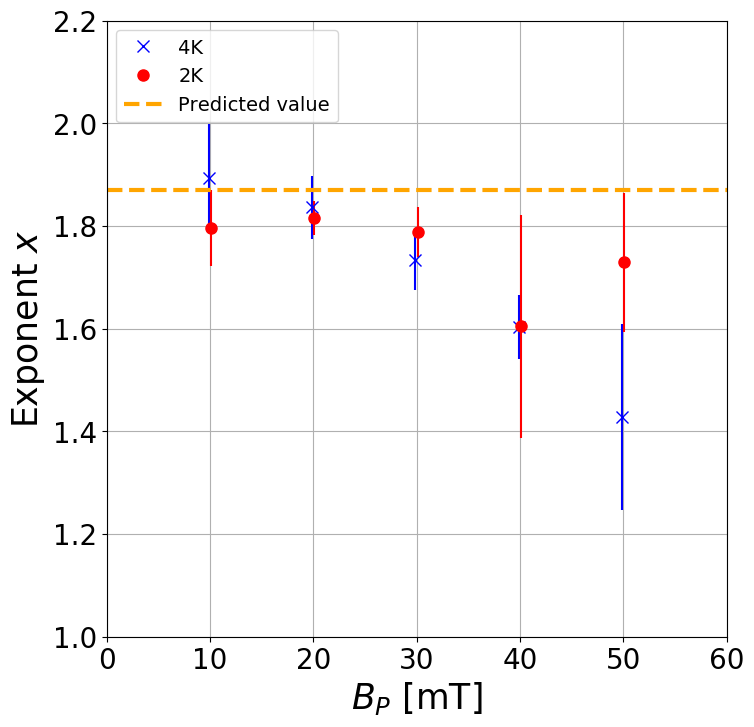}
\caption{\label{fig_RTdexponent} The exponent $x$ of the frequency dependence, Eq. (\ref{eq_FreqDependence}), matches the prediction, $x \approx 1.87$, by the theoretical model, Eq. (\ref{eq_RTd_theory}) (dashed line) within the determined uncertainties at low fields at both 4.2 (square markers) and 2\,K (circular markers), but trends towards lower values at higher field, deviating from the prediction.}
\end{figure}

Equation (\ref{eq_RTd_theory}) predicts $R_{Td}$ to rise with increasing frequency according to $\omega^{1.87}$. To determine the frequency dependence, $R_{Td}$ is plotted as function of frequency for fields of up to 50\,mT as shown in Figs. \ref{fig_RTd4Kfreq} and \ref{fig_RTd2Kfreq} for 4.2\,K and 2\,K respectively. Also shown are best fit lines in the form of  
\begin{align}
R_{Td}(\omega) = A_0 \omega^x \label{eq_FreqDependence}
\end{align}
with $A_0$ and $x$ as free fit parameters. Eq. (\ref{eq_FreqDependence}) will show up in the log-log plots as straight line with a slope equal to the exponent $x$. Based on the fitlines in Figs. \ref{fig_RTd4Kfreq} and \ref{fig_RTd2Kfreq}, the exponent $x$ seems to have a field dependence.  Fig. \ref{fig_RTdexponent} shows $x$ as a function of RF field for both temperatures. At low field and 4.2\,K, the exponent is determined as 1.9(1), which matches well with the predicted value of 1.87. At 2\,K and low field the exponent is lowered to 1.80(7), also matching the predicted value. While there seems to be a downward trend of $x$ with increasing RF amplitude, due to the fairly substantial uncertainty in the fits at higher fields or 2\,K it is difficult to determine any trend with certainty. Examination of this trend is subject to further studies.
\subsection{Temperature independent resistance}
\begin{figure}
\centering
\includegraphics[width=0.95\linewidth]{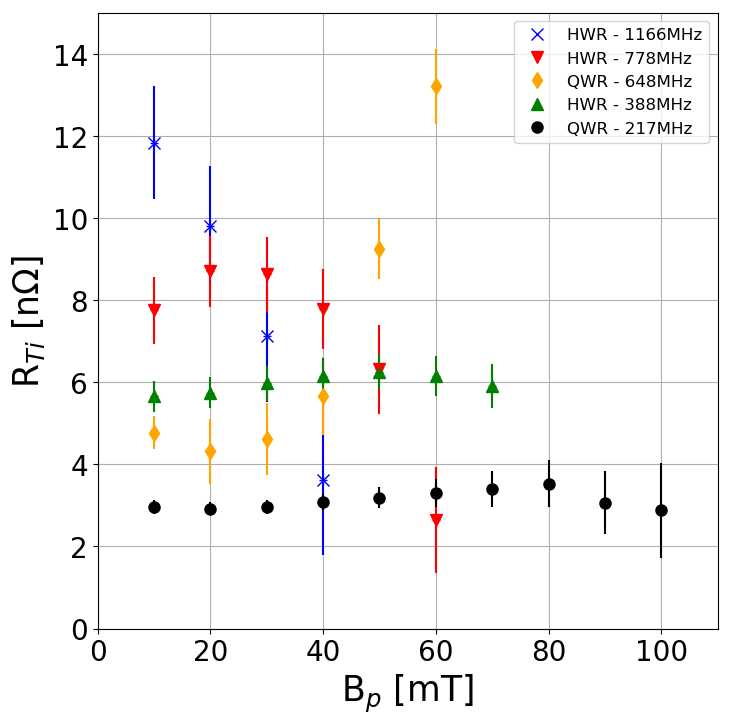}
\caption{\label{fig_RTiBp} Combined QWR and HWR $R_{Ti}(B_p)$ data for all investigated modes. No consistent trend between the modes can be determined. The sharp increase in the 648\,MHz modes at $B_p$ $>$ 40\,mT is attributed to of insufficient contaminant removal in a high field area of this mode.}
\end{figure}

\begin{figure}
\centering
\includegraphics[width=0.95\linewidth]{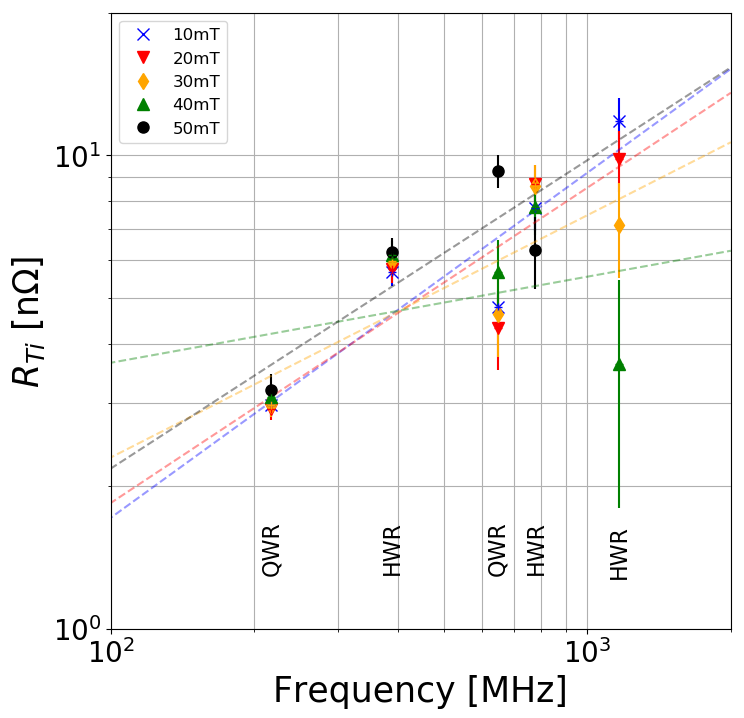}
\caption{\label{fig_RTiFreq} Combined QWR and HWR data for $R_{Ti}(\omega)$ reveals an increasing trend with increasing frequency for all field amplitudes, although there is a large scatter in the data. } 
\end{figure}

Figures \ref{fig_RTiBp} and \ref{fig_RTiFreq} show the field and frequency dependency of $R_{Ti}$ respectively. The sharp increase of $R_{Ti}$ at 648\,MHz at fields of 40\,mT and higher may be attributed to insufficient removal of contaminants after the heat treatment. Otherwise a fairly field independent trend is observed for the lower frequency modes, while a decrease in $R_{Ti}$ is observed for the high frequency modes. Regarding the frequency dependence, an overall increasing trend is extracted out of the cooldown data. Averaged over the measured RF field amplitudes, $R_{Ti}$ is $\propto \omega^{0.6}$. This is close to the frequency dependence of normal conducting losses in the anomalous limit of $\omega^{2/3}$. 
\subsection{QWR 120\degree\,C baking}
\begin{figure}
\centering
\includegraphics[width=0.95\linewidth]{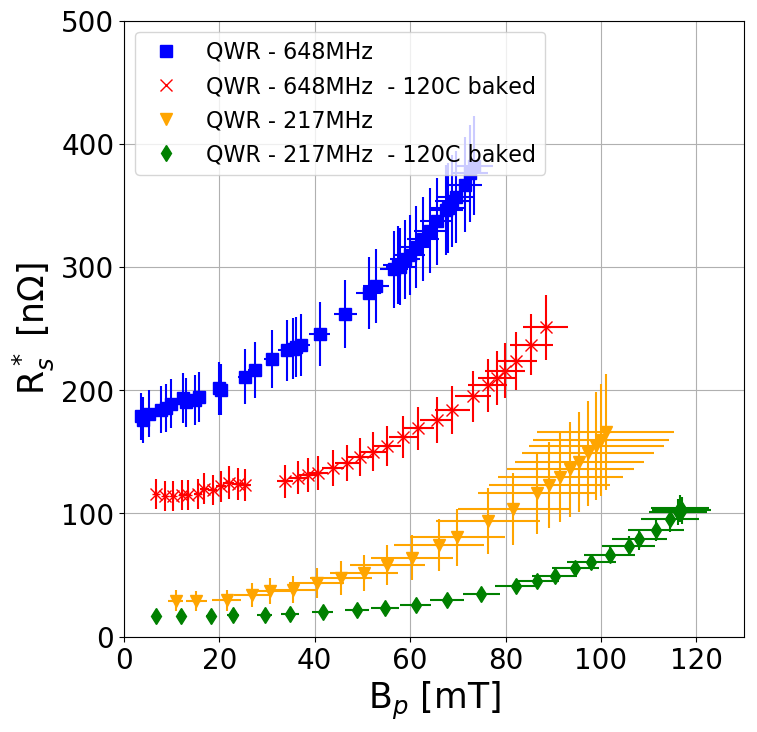}
\caption{\label{fig_RsBPbake4K} Baking the QWR at 120\degree\,C for 48\,h significantly reduces the uncorrected surface resistance $R_s^*$ at 4\,K in both modes.} 
\end{figure}

\begin{figure}
\centering
\includegraphics[width=0.95\linewidth]{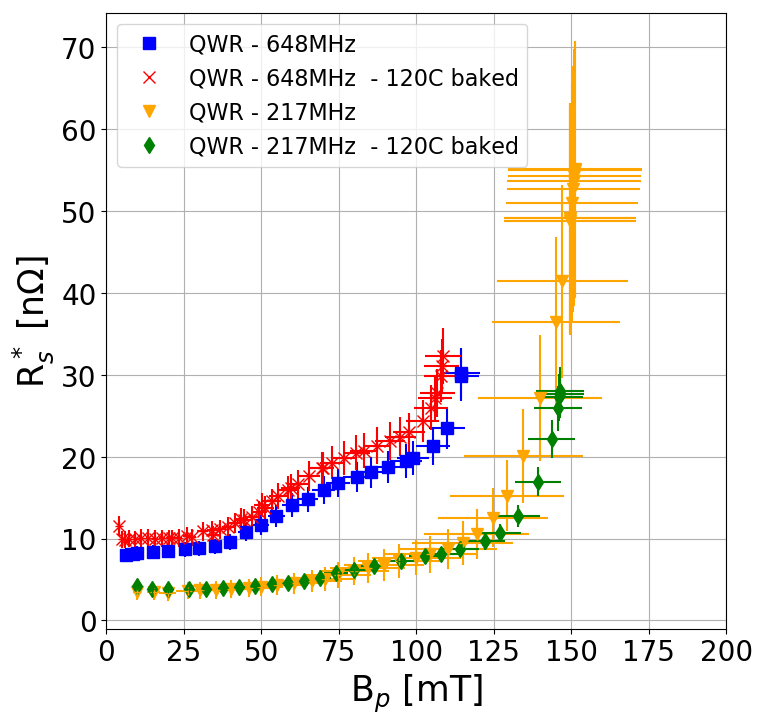}
\caption{\label{fig_RsBPbake2K} At 2\,K, the 120\degree\,C bake increases $R_s^*$ marginally for both QWR modes.} 
\end{figure}

A common cavity preparation is 120\degree\,C baking for 48\,h. In the presented case, the baking is done with resistive heaters strapped to the cavity, while the cavity is installed in the cryostat. During the bake both sides of the cavity wall, RF space and helium space surrounding the cavity, are under vacuum. The effect of this bake on $R_{s}^*$ of the QWR is shown in Figs. \ref{fig_RsBPbake4K} and \ref{fig_RsBPbake2K} for 4.2 and 2\,K respectively. A clear decrease in both amplitude and field dependence of $R_{s}^*$ is shown at 4.2\,K, while at 2\,K a slight increase in $R_{s}^*$ is visible. A conclusion can be made that the 120\degree\,C/48\,h treatment reduces $R_{Td}$, which dominates at 4.2\,K, while slightly increasing $R_{Ti}$. The reduction of $R_{Td}$ at 2\,K is insignificant compared to the increase of $R_{Ti}$. At the time of writing the HWR is in preparation for this surface treatment and once completed, a full analysis with frequency dependence will be done.

\subsection{Helmholtz Coil Demonstration}
\begin{figure}
\centering
\includegraphics[width=0.95\linewidth]{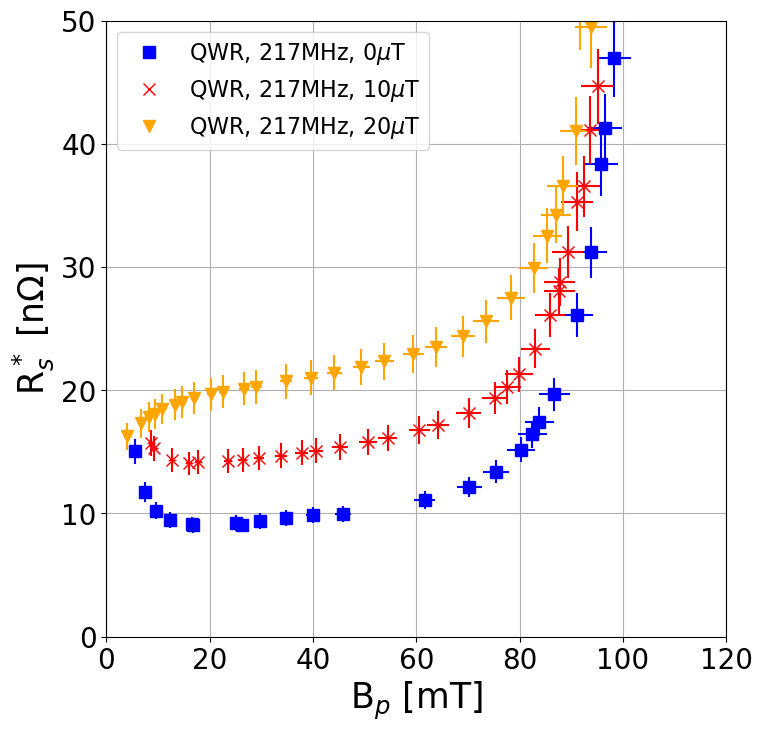}
\caption{\label{fig_Rmag217} $R_s^*$ data of the 217\,MHz mode measured at 2.1\,K with different dc external magnetic fields, aligned the the vertical cavity axis, shows a constant increase with the increased external field.} 
\end{figure}

\begin{figure}
\centering
\includegraphics[width=0.95\linewidth]{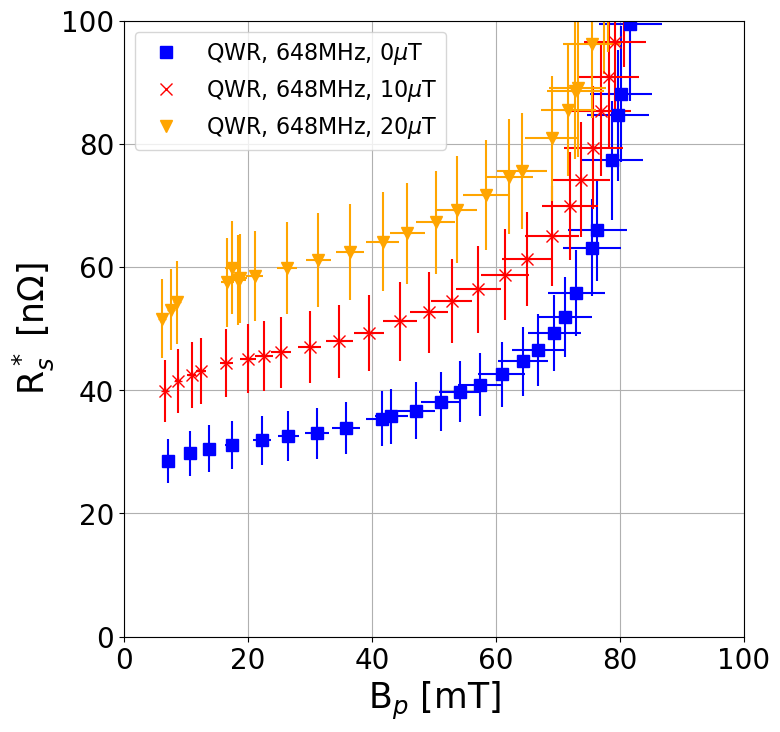}
\caption{\label{fig_Rmag648} At 648\,MHz a similar behaviour compared to the 217\,MHz mode is recorded when changing the external dc magnetic field between thermal cycles.} 
\end{figure}

The capabilities of the Helmholtz coils, were demonstrated with the QWR. The cavity was first cooled down in a fully compensated external field, with the current in all coils tuned to an external field in all three spacial dimensions of $<$0.5\,$\mu$T. After characterization, the cavity was warmed up above transition to around 20\,K, the vertical coils tuned to 10\,$\mu$T at the geometric center of both the cavity and coils, and then cooled down again below transition. This thermal cycle has been repeated with a field of 20\,$\mu$T. The resulting surface resistance  $R_S^*$ as function of peak surface field is shown in Figs. \ref{fig_Rmag217} and \ref{fig_Rmag648} for the 217\ and 648\,MHz modes of the QWR at around 2.1\,K. The slopes in the medium field range up to 80\,mT are identical between the different external fields but do have an offset to each other, suggesting a constant addition to the surface resistance caused by the external field. This amounts to a sensitivity $S$ of $\sim$0.5\,n$\Omega$/$\mu$T at 217\,MHz and $\sim$1.5\,n$\Omega$/$\mu$T at 648\,MHz.\\
In \cite{PadamKnoblochHays}, the magnetic field sensitivity $S$ is specified in Eq. (9.5) as 
\begin{align}
S = 3 \frac{\text{n}\Omega}{\mu\text{T}}\sqrt{f}
\end{align}
with $f$ as resonant frequency in GHz. Using this, a sensitivity of around 1.4 and 2.4\,n$\Omega/\mu$T at 217 and 648\,MHz respectively would be expected. The difference between textbook and measured value suggests that either not all the field is trapped in the cavity walls, or the flux is trapped in locations that do not contribute strongly to the surface resistance. Those would be areas with low magnetic surface fields, like the tip of the inner conductor, as these areas contribute to the losses significantly less than high field areas.\\ 
At 4.2\,K, the additional surface resistance is too small to be significant. Following similar measurements with the HWR, a full analysis including frequency dependence will be done.\\
Further studies are needed and planned to explore the role of trapped magnetic flux in TEM mode cavities and specific techniques to  mitigate reduced performance due to flux trapping. 
\section{Summary\label{secConclusions}}
The TRIUMF multi-mode coaxial SRF cavities are an excellent tool to study TEM mode cavities. In particular, the dependence of the surface resistance on temperature, surface treatment, RF frequency, external magnetic field, and RF field amplitude are available to study, opening an unprecedented parameter space to be explored. The presented infrastructure in place at TRIUMF allows for exploration of this large parameter space. RF amplitudes of up to 150\,mT peak surface magnetic field have been reached. In the presented data, some early conclusions are drawn on the field and frequency dependence. \\
Characterization of both the QWR and HWR after degassing at 800\degree\,C and a flash BCP surface removal show excellent performance both at 4.2\,K and 2\,K, on par with performances of 1.3\,GHz single cell elliptical cavities with the same surface treatment. The cavities show a low surface resistance and high quench field. \\ 
Data collected during the cooldown at several RF field amplitudes and multiple resonant modes allows to separate $R_s$ into its components $R_{Td}(T, B_p,\omega)$ and $R_{Ti}(B_p,\omega)$ and analyse the frequency and field dependence of these parameters. \\
The data reveals that the temperature dependent term $R_{Td}$ at low RF fields is $\propto \omega^{1.9(1)}$ at 4.2\,K and $\propto \omega^{1.80(7)}$ at 2\,K, matching with the predicted dependence of $\omega^{1.87}$. The RF field dependence of $R_{Td}$ matches both a quadratic and an exponential growth model in the investigated range of field amplitudes. The change in slope between 4.2 and 2\,K is dominated by the reduction of the zero field resistance $R_0$, rather than the slope parameter $\gamma$. \\
The temperature independent component $R_{Ti}$ gives a less clear picture due to a large scatter in the data. An overall increasing trend with increasing frequency $\propto \omega^{\sim 0.6}$ is found, which matches with anomalous losses. No clear conclusion can be drawn on the RF field dependence.\\
Capabilities to bake the QWR at 120\degree\,C have also been demonstrated, which resulted in a significant higher $Q_0$ at 4.2\,K, and a small decrease in $Q_0$ at 2\,K. This is attributed to a strong decrease in $R_{Td}$, which is the dominant term at 4.2\,K, and a small increase in $R_{Ti}$, which is of comparable order to $R_{Td}$ at 2\,K.\\
The functionality of the Helmholtz-coils has been demonstrated and a first estimation of the external magnetic field sensitivity for a vertical field orientation measures a sensitivity of the QWR of $\simeq$0.5\,n$\Omega$/$\mu$T at 217\,MHz and $\simeq$1.5\,n$\Omega$/$\mu$T at 648\,MHz.
\subsection*{Future Work}
This paper shows the possibilities of the research areas covered by the coaxial multi-mode cavities at TRIUMF with examples of early performance measurement results. Future work will include comprehensive studies of the effects of various surface treatments, as well as changes of the background magnetic field, on the surface resistance. A further step in data preparation will also include corrections between the measured helium bath temperature and the RF surface temperature of the cavity, as well as measurements at lower temperatures to fully observe the expected levelling off of the surface resistance at low temperatures. Further planned infrastructure improvements are electro-polishing for surface treatments, and a temperature mapping system to further advance understanding of the details of the surface resistance.\\
\section*{Acknowledgements}
The authors would like to thank the technical team of the SRF group, namely of Devon Lang, James Kier, Ben Matheson, and Bhalwinder Waraich, and cryogenics group, namely Johnson Cheung and David Kishi, who helped tremendously with the collecting the data. Funding is provided by NSERC.
\bibliography{References}

\end{document}